# RF Beam Steering Using On-Chip Optical True Time Delay on VTT's 3 μm SOI Platform

SOMNATH PAUL,† JUSSI SÄILY, MARKUS SCHRÖDER, MIKKO HARJANNE AND TIMO AALTO

*VTT Technical Research Centre of Finland, Tietotie 3, 02150 Espoo, Finland*
**somnath.paul@vtt.fi*

**Abstract:** We present the design, packaging, and system-level experimental demonstration of a reconfigurable 5-bit, 4-channel optical true time delay (OTTD) chip realized on VTT's 3 μm thick silicon-on-insulator (SOI) platform. The chip enables 32 discrete delay states over a 0–101 ps range, with a delay resolution of ~3.25 ps, making it well-suited for broadband RF beamforming with upcoming 6G RF3 band systems around 10 GHz. Each channel incorporates five binary-weighted delay stages controlled by Mach–Zehnder interferometer (MZI)-based 2×2 thermo-optic switches with an extinction ratio exceeding 25 dB and an average switching power of ~25 mW. The chip is embedded within an analog radio-over-fiber (A-RoF) link, and the outputs are connected to a PCB-based four-element patch antenna array. Accurate RF beam steering is demonstrated at broadside and at 30° from broadside using a 16-QAM signal modulated at 100MBaud on a 10 GHz carrier. An error vector magnitude (EVM) degradation of approximately 1% is observed upon steering, and an extinction ratio exceeding 50 dB is measured between the two receiver positions. The results confirm the potential of integrated OTTD technology for compact, power-efficient, and broadband beamforming in next-generation wireless and radar systems.

## 1. Introduction

Analog RF beam steering is a key component of next-generation wireless communication systems, especially at frequencies above 7 GHz, where the cost and power consumption of high-resolution ADCs and DACs make pure digital beam steering impractical [1,2]. Conventional analog RF beam steering relies on electronic phase shifters within phased array antennas (PAA), where the beam direction is controlled by the relative phase difference between array elements. These electronic phase shifters are bulky (passive elements), power hungry (active elements) and inherently frequency dependent [3]. This high frequency dependence, in the full RF chain, causes significant degradation of system performance during broadband operations due to the well-known beam squint effect where the beam direction shifts with signal frequency. Optical true time delay (OTTD) offers a promising alternative to electronic phase shifters by delaying the RF signal in optical domain, making it independent of RF bandwidth and overcoming the beam squint effect [4,5]. Furthermore, the integrated on-chip OTTDs brings additional advantages in terms of cost, footprint and power consumption.

With the advancement of photonics integrated circuit technology, a variety of different schemes such as ring resonator based [6], Mach Zehnder Interferometer (MZI) switch based [7], photonic crystal based [8], waveguide Bragg grating based [9] OTTDs have been implemented in the past with different material platforms [10–12]. These works have demonstrated varying degrees of delay range, resolution, insertion loss, and system complexity. However, comprehensive end-to-end system demonstrations that include actual RF beam steering with baseband-modulated signals at microwave carrier frequencies remain relatively scarce, particularly for chips fabricated on thick SOI platforms which offer inherently lower propagation losses than thin SOI counterparts.

VTT's 3 μm thick SOI platform [13] is distinguished by its exceptionally low propagation loss of approximately 0.03 dB/cm at 1.55 μm optical wavelength [14]. The micron-scale dimension of rectangular waveguides shows very low polarization dependency in this platform. This low loss and near-zero birefringence make it particularly attractive for realizing long delay lines without incurring prohibitive insertion loss and polarization scrambling which is a key requirement for wideband OTTD applications [15].

In this work, we present a detailed account of the design methodology, chip architecture, packaging strategy, system integration, and experimental characterization of a 5-bit, 4-channel OTTD chip fabricated on VTT's 3 μm SOI platform. This chip is used to demonstrate RF beam steering in a system-level setup incorporating an analog radio-over-fiber (A-RoF) link and a four-element RF patch antenna array operating at 10 GHz. The paper is organized as follows: Section 2 discusses the design of the OTTD chip, including the delay line architecture and MZI switch design. Section 3 describes the electrical and optical packaging of the OTTD chip including an interposer scheme to facilitate better mode matching between the fiber and chip edge. Section 4 presents the system-level architecture and components of the experimental setup. Section 5 reports experimental results, including switch characterization, delay verification, and beam steering measurements. Section 6 summarizes the paper with discussion.

## 2. OTTD Chip design

### *2.1 Delay line architecture*

The OTTD chip is designed for a 4-element phased array antenna system with four identical optical channels, each of which contains 5 identical delay stages. The operating carrier frequency is chosen as 10 GHz which falls somewhat at the center of the possible future 6G FR3 band (7 – 15 GHz) as predicted by industry leaders [16,17]. In time-domain, the 10 GHz carrier RF frequency has a period of 100 ps. The 5-bit reconfigurable delay architecture provides 32 discrete and uniformly spaced delay values between 0 and 100 ps covering one full cycle of the carrier signal with delay accuracy of 3.225 ps. The adjacent delay stages differ by a factor of two, following a binary-weighted progression from the least significant bit (LSB) to the most significant bit (MSB). The delay lengths of the five stages are ΔL, 2ΔL, 4ΔL, 8ΔL, and 16ΔL, respectively, where ΔL is the unit delay increment as shown schematically in Fig. 1(a). With this configuration, the maximum delay achievable on a single channel is 31ΔL, covering all 32 states from 0 to 31ΔL. The required delay length for each stage is determined by the optical propagation delay formula:

$$t_{\Delta\mathrm{L}} = \frac{delay\ length \times n_g}{c} \tag{1}$$

where $t_{\Delta\mathrm{L}}$ is the desired time delay, $n_g$ is the group index inside the delay waveguide, and $c$ is the speed of light in vacuum.

All delay lines are formed with strip waveguides having a horizontal dimension of 1.5 μm and a vertical dimension of 3 μm, with the group index for the fundamental TE mode at 1.55 μm optical wavelength $n_g$ = 3.6963. Hence, from Eqn. (1), the total delay length for a maximum of 100 ps time delay is 8.122 mm with unit length increment of ΔL = 0.262 mm. It is important to note that within each stage, the light traverses either the delay arm or a reference arm of equal construction. The effective delay introduced by each stage is the difference in optical path length between the delay and the reference arms. Consequently, the "zero" delay state for a given channel corresponds to the sum of all reference arm lengths, and only the incremental differences determine the actual time delays. The input optical power is distributed equally across all four channels by means of two cascaded stages of 1×2 multimode interference (MMI) couplers. This binary tree splitting structure ensures uniform power delivery to each channel, which is essential for maintaining a consistent signal-to-noise ratio across all the antenna elements. A picture of the fabricated OTTD chip after dicing is shown in Fig. 1(b).

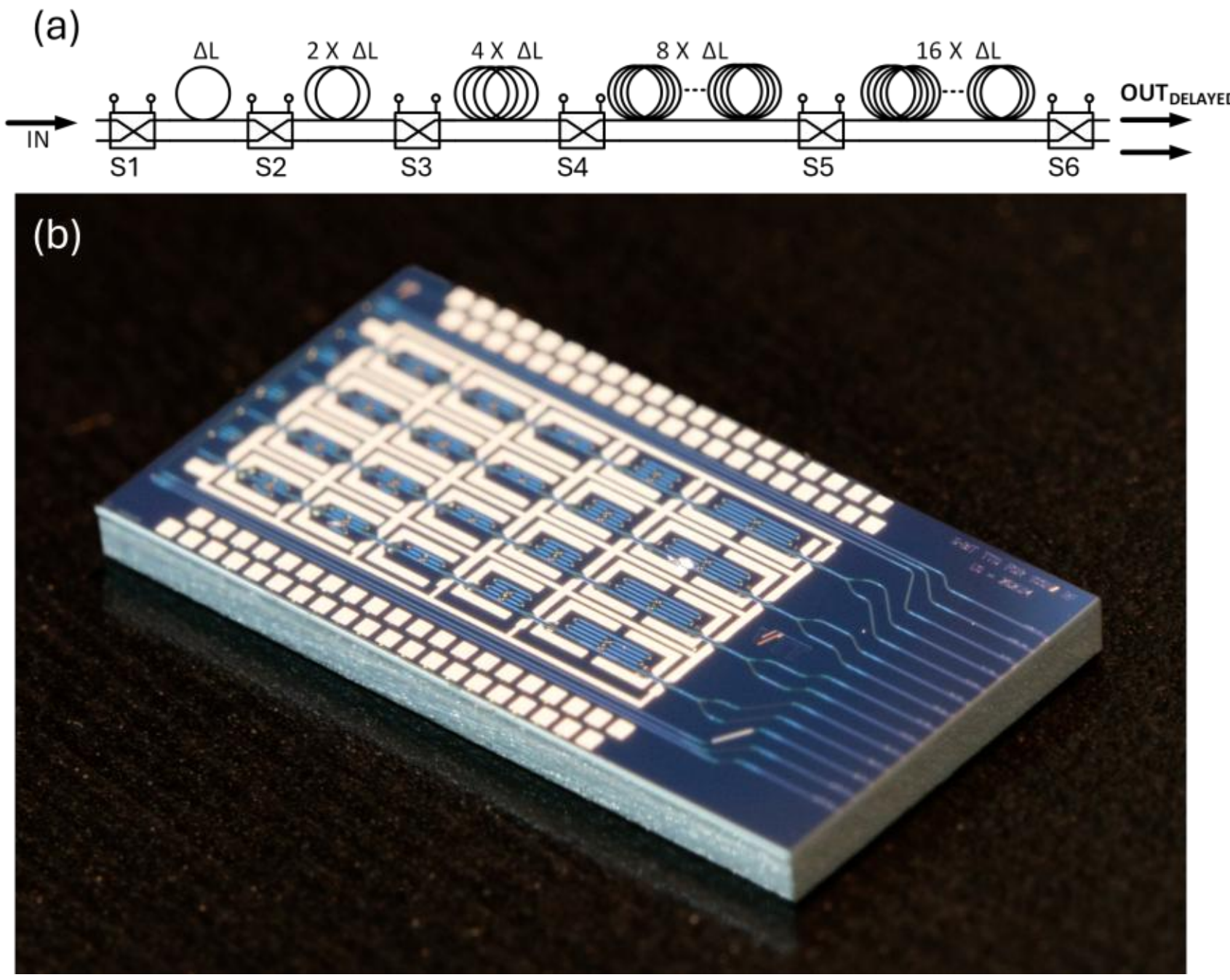

Fig. 1. (a) The schematic of a 5-bit delay line, (b) Picture of the diced OTTD chip.

### *2.2 MZI switch design*

The delay stages are separated by Mach-Zehnder interferometer (MZI)-based $2 \times 2$ switches. Each channel has 6 MZI switches with first 5 serve the 5 delay stages, and the 6$^{th}$ routes the output. The "Cross" or "Bar" state of a switch decides if the light passes through the delay arm of the following stage or the reference arm of that stage. Each MZI switch consists of two identical $2 \times 2$ multimode interference (MMI) couplers connected by two waveguide arms with in-line thermo-optic (TO) phase shifters. The MMI couplers are designed to provide a 50/50 splitting ratio, functioning as balanced 3 dB couplers for the MZI at 1.55 µm wavelength. For the switching operation both the phase shifters of a MZI can be used simultaneously in push-pull configuration to utilize the maximum switching frequency.

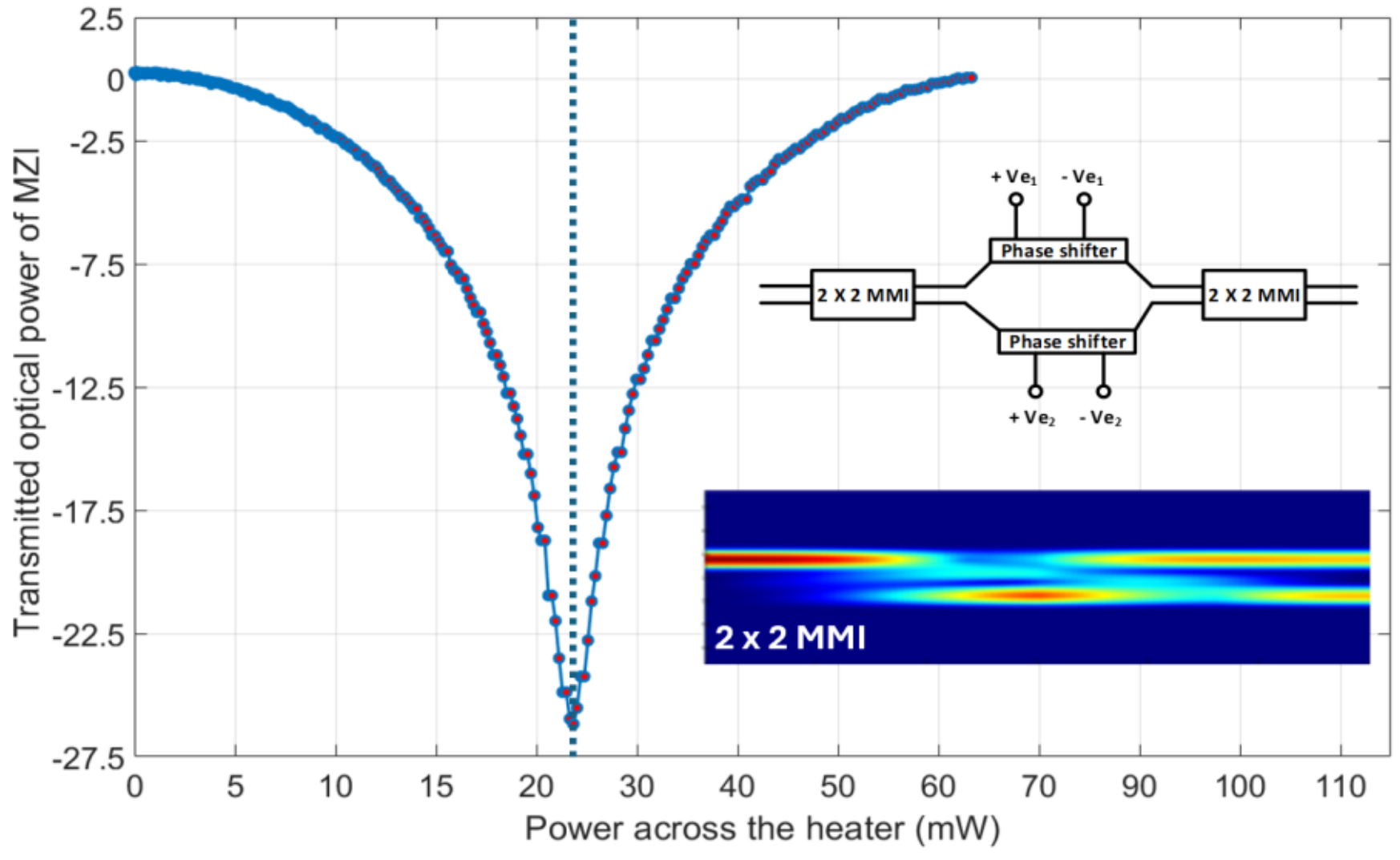

Fig. 2: Transmission characteristics of the MZI with applied power.

The measured extinction ratio between the cross and bar states of the MZI switch exceeds 25 dB, ensuring high-fidelity switching with minimal crosstalk between the delay and reference arms. The average switching power defined as the electrical power required to induce a $\pi$-phase shift is approximately $P_{\pi}$ = 25 mW in VTT's 3 µm SOI platform. The optical transmission characteristic of the switch is shown in Fig. 2. A schematic of the 2× 2 MZI switch with the field distribution of TE polarized light at 1.55 µm for balanced MMI coupler is shown in the inset of Fig. 2. For applications where switching speed is not a critical requirement, the switching power can be further reduced to 2–3 mW by placing dedicated thermal isolation cavities beneath the TO phase shifters [18].

## 3. Optical and electrical packaging

The fabricated OTTD chip is edge coupled with 250 µm pitch single mode fiber (SMF) array via a 12 µm photonic interposer. The interposer adiabatically tapers from 12 µm SMF mode field to 3 µm device layer mode field of the OTTD chip to ensure efficient fiber-to-chip optical power coupling, as explained in detail in [18]. The TO phase shifters on the OTTD chip are terminated with metallic contact pads. The chip is placed on a PCB-based host board and wire bonded from the PCB. The host PCB routes the electrical connections to a standard connector interface to get access from a multi-channel programmable DC power supply. A photograph of the fully packaged OTTD chip, showing the fiber array and electrical interface is presented in Fig. 3.

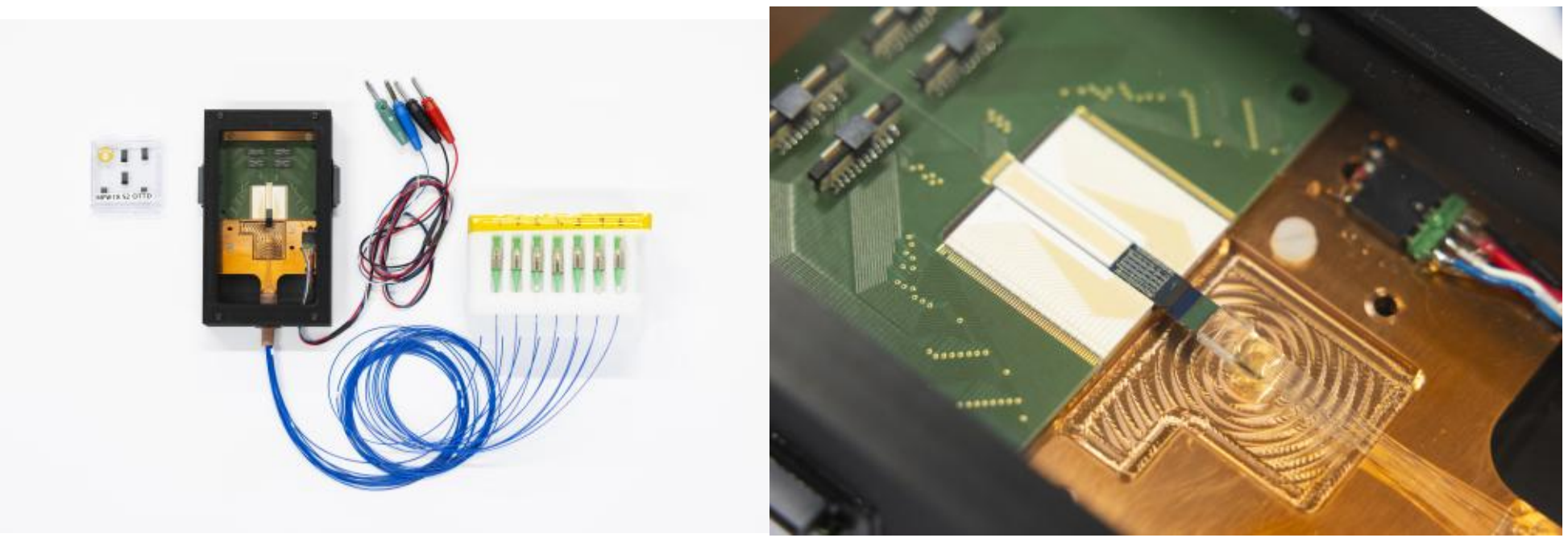

Fig. 3: Packaged OTTD chip on the left with fiber connections and electrical contacts. Zoomed-in part of the OTTD chip on the host board on the right.

## 4. System architecture and experimental setup

### 4.1 Analog radio-over-fiber link

To evaluate the system-level performance, the OTTD chip is placed in-line with an analog radio-over-fiber (A-RoF) link, as illustrated in Fig. 4. The A-RoF link consists of a continuous wave (CW) laser source of optical wavelength 1550 nm and 100 kHz linewidth, a electro-optic Mach-Zehnder modulator (SilOrix GmbH) to upconvert the RF signal to optical domain, an RF signal generator with 10 GHz carrier and 16-QAM modulation capability at 100 MBaud, and fast photo detectors to retrieve the RF signal. Note that here the QAM modulation is in the RF domain that modulates only the intensity of the laser at the RF carrier frequency with the help of an Mach-Zehnder modulator (MZM), and both I (in-phase) and Q (quadrature) components of the RF-QAM are recovered with the fast photo detector. Four identical replicas of RF modulated optical signal passes through the 4-channel OTTD chip and terminates at four identical photo diodes but with four independent time delays.

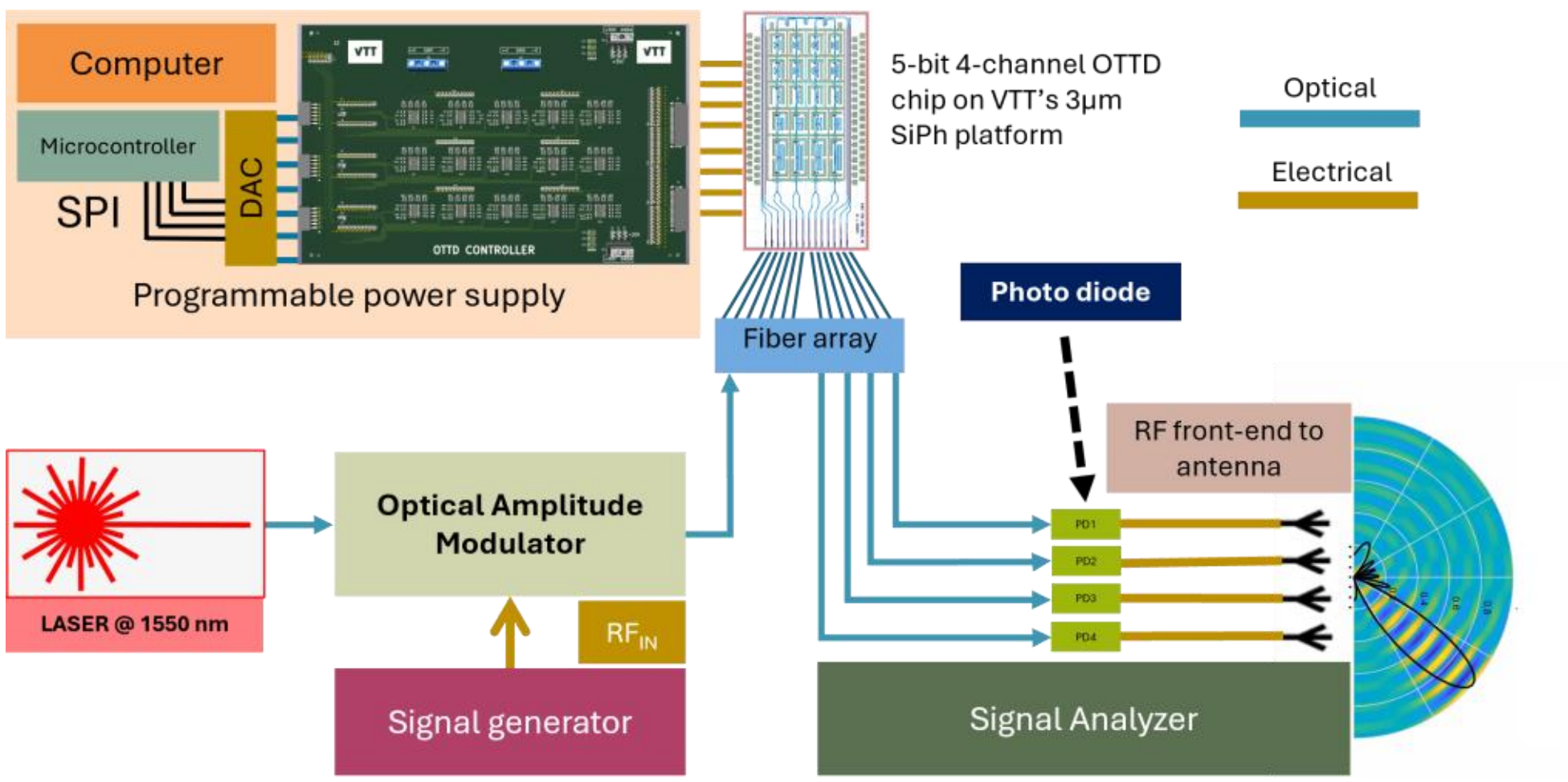

Fig. 4: Block diagram of the system-level setup for the OTTD-based RF beam steering.

### *4.2 Switch control and calibration*

The 5-bit, 4-channel OTTD chip consists of 48 MZI switches, each of which is connected to a programmable power supply. The power level of each supply channel is adjusted with 12-bit digital-to-analog converters (DACs) providing 4096 discrete levels. The range of the power supply is scaled with external buffer electronics to ensure most of the DAC levels fall within 0 and $P_\pi$ to tune the MZI switches at their optimal bar and cross states. An automated algorithm is used to identify the optimal states of each switch in a non-invasive manner similar to the method described in [20] and stored in memory for future use.

Following the individual switch characterization, the chip was further calibrated to construct a code book to define the logical "0" and "1" state of each delay stage, and for all the four channels of the OTTD. This code book maps each of the 32 desired delay states (for each of the four channels) to a specific bitstream (0 = zero power and 1 = $P_\pi$) applied to the six switches in that channel. For example (see in Fig. 1(a)), a bitstream 101010 corresponds to S1 = $P_\pi$, S2 = 0, S3 = $P_\pi$, S4 = 0, S5 = $P_\pi$, and S6 = 0, respectively. The code book considers the variation of optimal operating point of individual switches while defining the bitstream for a certain delay using the pre-stored data from characterization round.

### *4.3 Antenna array and receiver setup*

The recovered RF signals after the fast photo detector for all the four channels are fed into a PCB based patch antenna designed to operate at 10 GHz carrier frequency. Identical gain amplifiers are used between the photo diode and antenna to ensure sufficient beam power. The antenna elements are arranged in a linear array configuration with a half-wavelength ($\lambda_{RF}$ = 3 cm) inter-element spacing, consistent with standard phased array design practice.

Two identical horn antennas are used as receivers. One is positioned at broadside (0° with respect to the transmitter) and the other at a fixed angle of 30° from broadside. Note that the RF beam steering range is only limited by the antenna array design, not by the number of channels. Both receivers are placed 5 meters from the center of the transmitting phased array. The signals received from both horn antennas are simultaneously acquired on two separate channels of a high bandwidth signal analyzer, enabling concurrent measurement of the signal quality at broadside and at the steered direction. Two identical low noise amplifiers (LNA) are used between the receiver antenna and signal analyzer. Note that the LNAs also amplify the noise coming with the free-space signal and a bandpass channel filter would be good to have on the receiver side to receive a cleaner signal. The images of the patch and horn antennas used in this experiment are shown in Fig. 5 with current distributions and simulated beam patterns.

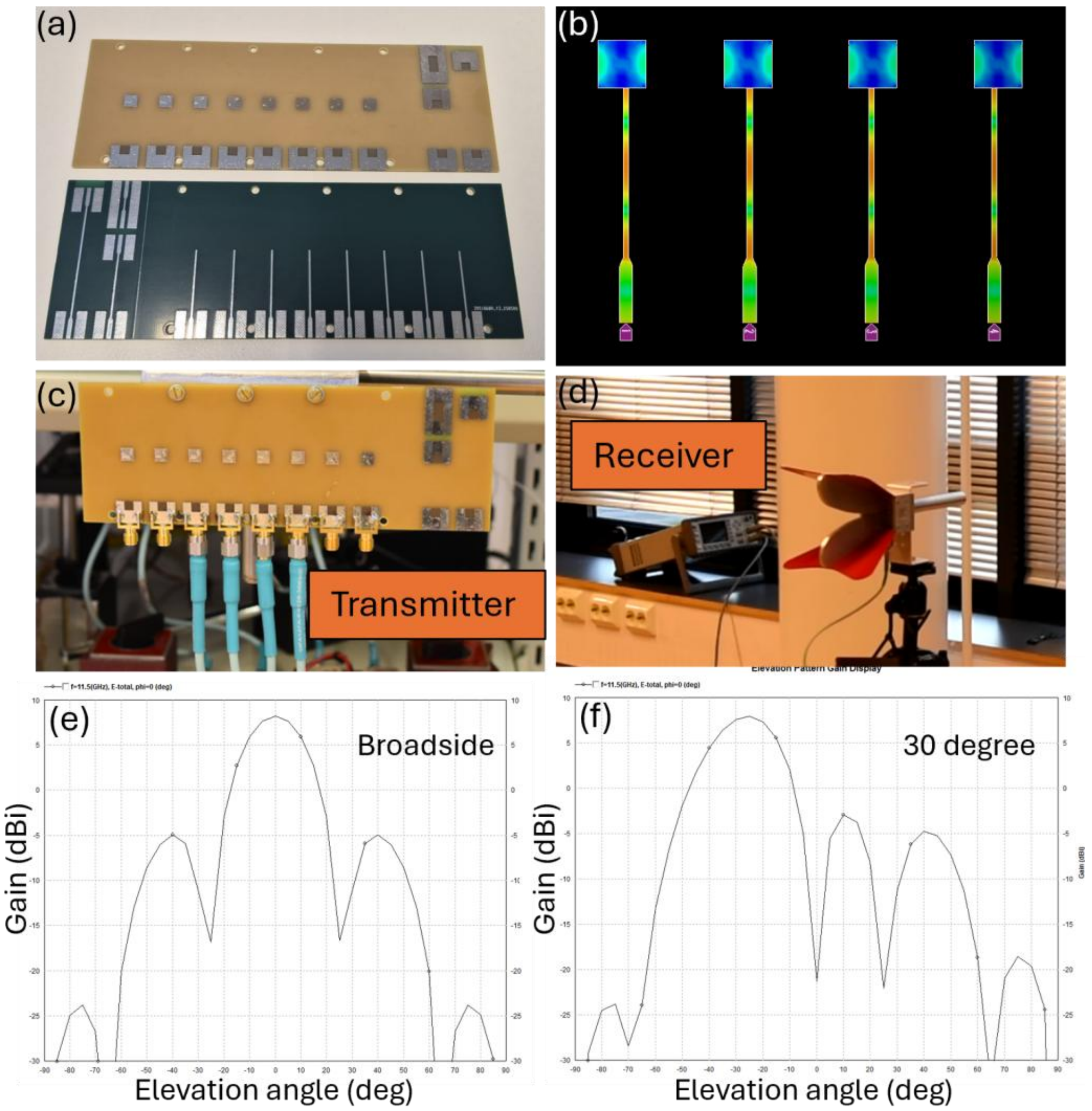

Fig. 5: (a) Picture of the PCB-based patch antenna both front and back side, (b) the current distribution for 10 GHz RF carrier frequency, (c) front side of the patch antenna with RF cable connections, and (d) the receiver horn antenna. (e) and (f) are the calculated beam shape in 2D for broadside and 30-degree beam, respectively.

## 5. Experimental results

### *5.1 Insertion loss characterization*

The optical loss in the OTTD chip is measured for different delay configurations. For the maximum delay, where light is routed through all the delay arms, the insertion loss is 1.8 dB per channel. This loss value is compared with the reference delay and few intermediate delay states, and no significant difference is observed. This measurement confirms that the delay lines themselves make negligible contribution to the insertion loss which is expected for 3 µm SOI platform (loss < 0.03 dB/cm) and a maximum delay of 8.122 mm. The total loss of 1.8 dB per channel is coming from the cumulative contributions of 12 MMI couplers (0.15 dB loss per

MMI) of 6 MZI switches across the delay stages. The fiber to chip coupling loss, measured with reference waveguide on the same chip, is 2.2 dB per facet.

### *5.2 Delay characterization*

The delay performance of the OTTD chip is performed by launching optical signal modulated with 10 GHz continuous wave RF signal. The four independent photo detector outputs, after the 4-channel OTTD chip, are connected to a 4-channel fast oscilloscope. All the 31 relative time delays of each channel are measured from the oscilloscope time scale one after another while keeping the rest of the channels fixed in phase as reference. As an example, in Fig 6, the oscilloscope waveforms for four different delay configurations for channel 2 (Ch2) are presented. In configuration (a), Ch2 is phase aligned with the rest of the channels corresponding to zero-delay. In configuration (b), (c), and (d), Ch2 shifts in time scale from the reference position by 13 ps, 26 ps and 52.1 ps, respectively. The measured total delay is 101 ps with a delay resolution of ~ 3.258 ps.

We have calculated the effect of defining the beam direction with the ideal delay and the delay that we could apply due to discretization and deviation of delay resolution from the target design to the four channels for a beam angle of 30 degrees. The required progressive time delay between the antenna elements is calculated from standard array factor relation for a linear phased array system [21]. To steer the beam at 30° relative to the broadside, the required delays are Ch1 = 0 ps, Ch2 = 25 ps, Ch3 = 50 ps and Ch4 = 75 ps. We set the nearest possible delays to the corresponding channels and calculated the change in beam shape and angular deviation ($\varphi_{err} = 0.3^{\circ}$) as shown in Fig. 6(e) with array factor calculation. A comparative table of ideal vs applied delays are presented in Fig 6(f).

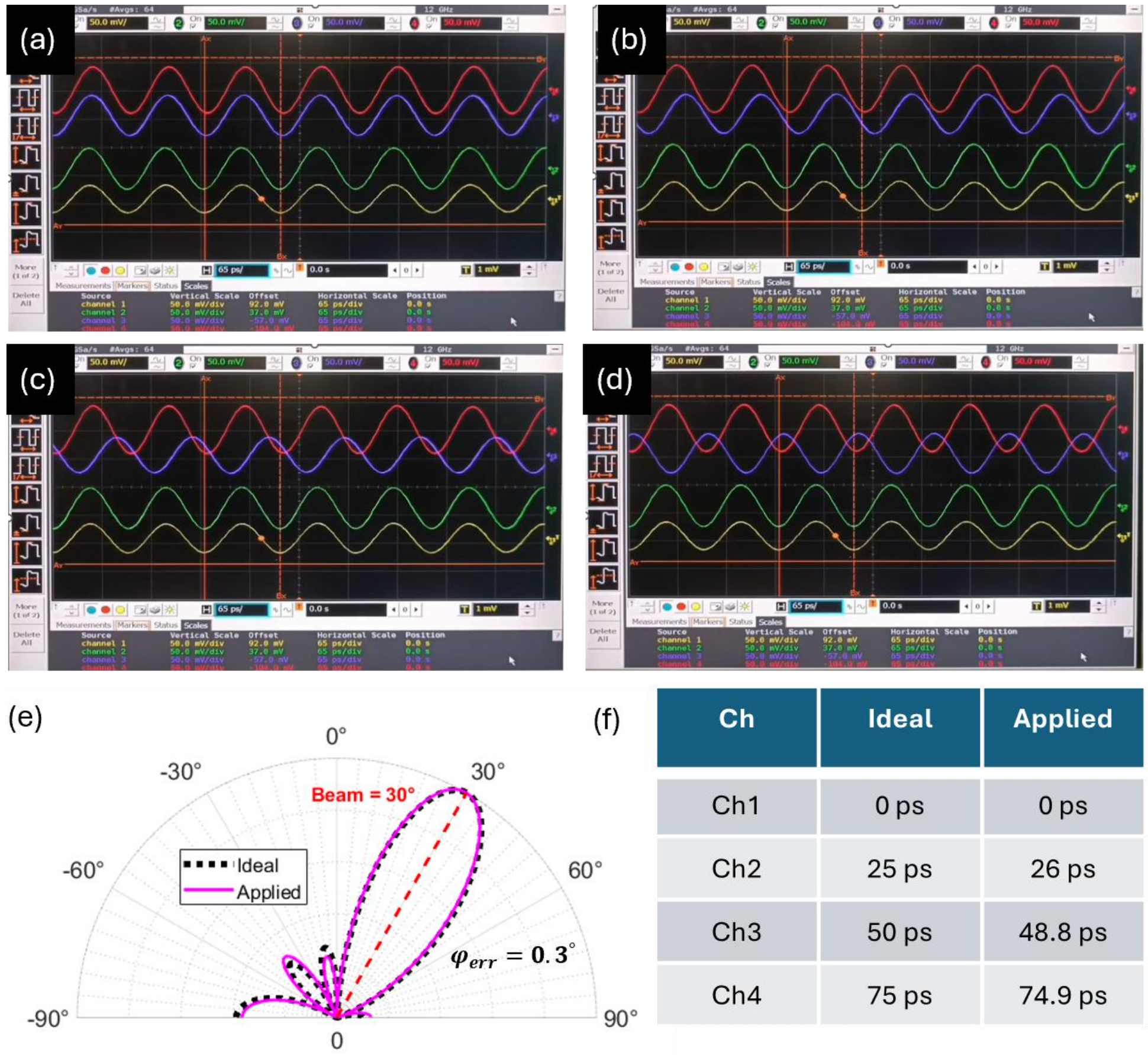

| Ch | Ideal | Applied |
| --- | --- | --- |
| Ch1 | 0 ps | 0 ps |
| Ch2 | 25 ps | 26 ps |
| Ch3 | 50 ps | 48.8 ps |
| Ch4 | 75 ps | 74.9 ps |

Fig. 6. (a) – (d) Relative position of Ch2 (blue) relative to Ch1, Ch3, and Ch4. (e) Comparison of the beams formed with ideal vs applied delay as presented in the table in (f).

### *5.3 RF beam steering demonstration*

The system-level evaluation of the RF beam steering is performed by feeding the photodetector outputs into a 4-channel transmitting patch antenna. Four identical gain amplifiers are used to increase the signal level for sufficient free space propagation. An arbitrary waveform generator (Keysight M8196A) is used to generate 10 GHz RF carrier with16-QAM baseband data at 100 MBaud and fed to the modulator in the A-RoF line (see Fig. 4). At first, the OTTD chip is configured utilizing the pre-calibrated code book to construct the RF beams to the broadside (0°).

For the 4-element patch antenna operating at 10 GHz the calculated delays are 25 ps, 50 ps, and 75 ps to steer the RF beam at 30°. However, due to the limited delay resolution, Ch2, Ch3, and Ch4 of the OTTD chip is configured to have delays of 26 ps, 48.8 ps, and 74.9 ps, respectively. The received signal from both the receiver antennas is fed simultaneously into a fast oscilloscope (Keysight UXR0334B) in two different channels. The observed error vector magnitude (EVM) for broadside receiver and 30° receiver is 7.8 % and 9.1 %, respectively. The absolute power level extinction between steered beam directions is over 50 dB, suggesting efficient beam steering with the OTTD chip with good signal integrity.

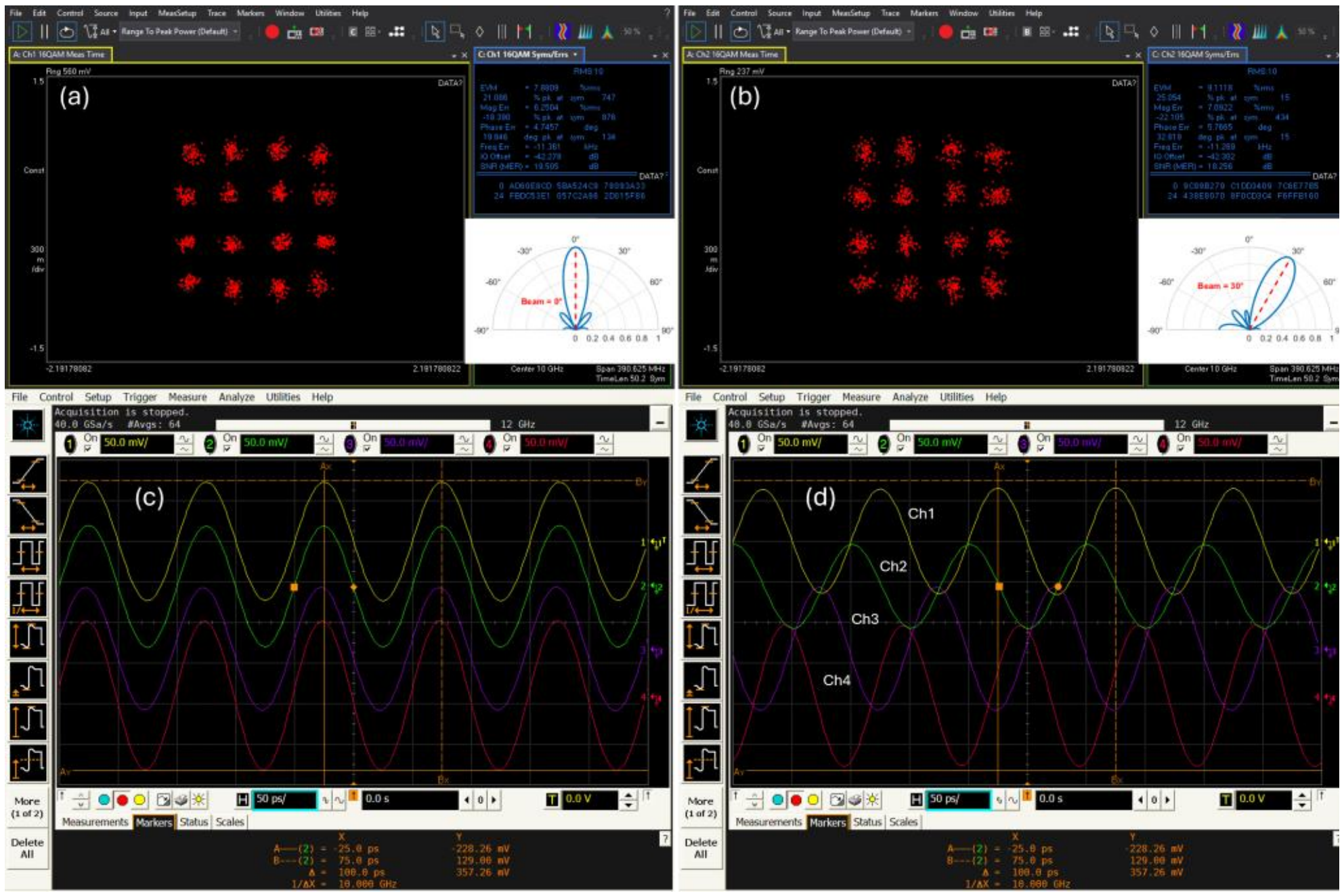

Fig. 7. Comparison of 16-QAM constellation for the received signal at (a) broadside, and (b) 30° beam. Oscilloscope waveform indicating the relative time delay between the transmitter antenna elements (c) for broadside and (d) for 30° beam. See Visualization 1.

## 6. Discussion

This work presents a complete system-level validation of RF beam steering using a 16-QAM 100 Mbaud baseband signal with a physical RF frontend, establishing the practical viability and high integration level of OTTD-based systems. Beam steering accuracy can be further enhanced by increasing the number of delay stages per channel, thereby expanding the bit count. The low insertion loss of approximately 0.3 dB per delay stage in VTT's 3 μm SOI platform makes it particularly well-suited for realizing OTTDs with a larger number of bits. The combination of low propagation loss (0.03 dB/cm), high mode confinement, and large mode volume renders VTT's 3 μm SOI platform exceptionally well-suited for the realization of long delay lines within a compact footprint, while supporting optical power levels up to 1 W. This high-power handling capability opens up the possibility of feeding a larger number of channels from a single optical input, thereby enhancing the scalability of the system. In contrast, thin SOI (220 nm) platforms typically exhibit significantly higher propagation losses, constraining the number of achievable delay stages and the total delay range before insertion loss becomes prohibitive. Furthermore, the inherently small mode volume of thin SOI platforms limits the maximum tolerable optical power, as the high optical intensity promotes nonlinear absorption effects. While silicon nitride platforms offer propagation loss values comparable to the 3 μm SOI platform, their lower index contrast requires larger waveguide bends and longer delay paths, making compact, high-bit-count integration more challenging. Table 1 compares the features of our OTTD chip with other reported OTTD approaches, including switched-delay-line, ring-resonator, photonic-crystal, and Bragg-grating designs.

Table 1. Comparison of different representative integrated OTTD approaches

| Reference | Platform / Mechanism | Delay line specification | Waveguide Loss | Overall loss / ps | System-level demo with RF front end |
|---|---|---|---|---|---|
| This work | 3 μm SOI, MZI switch | 5-bit discreet | 0.03 dB/cm | 0.015 dB | Yes |
| Martinez-Carrasco et al. [4] | 220 nm SOI, MZI switch | 5-bit discreet | 1.1 dB/cm | 0.026 dB | No |
| Tsokos et al. [5] | LioniX, TriPleX, Si3N4, MMR | Continuous | 0.1 dB/cm | Not reported explicitly | Yes |
| Sancho et al. [8] | InP photonic-crystal slow-light waveguide | Continuous | Not reported explicitly | 0.14 dB | No |
| Burla et al. [9] | SOI dual-phase-shifted waveguide Bragg grating | Continuous | Not reported explicitly | 0.25 dB | NA |

In this demonstration, the beam steering angle of 30° is considered due to the limitation of beam steering to larger angles than 35°. This limitation arises from the chosen simple patch antenna design having a half-power beamwidth around 70°. Scan loss (difference in antenna gain between boresight and scanned directions) will quickly increase beyond the expected 3 dB toward 35°. The expected scan loss toward any direction of a beam steered antenna array follows the radiation pattern envelope of a single unit cell antenna in the array. Thus, to increase the scanning capability, a wide beamwidth antenna element such as a dipole antenna should be used in the antenna array. Modern base station antenna arrays are designed to work in 120° sectors and can have ±60° scanning capability within the sector.

Note that the delay range (0 ps to 100 ps) of the OTTD chip is sufficient to form the beam with larger angles for 10 GHz carrier RF signal by rolling over the delay values after 100 ps, hence, there is no need for progressive time delay. For massive-MIMO application, the demonstrated architecture with 4-channels can be directly scaled for higher number of channels by externally cascading multiple 4-channel OTTDs with 1×N fiber-based beam splitters.

The TO phase shifters in VTT's 3 μm SOI platform is practically loss less, however, the switching speed is around sub-millisecond range. For applications, where the beam scanning frequency is higher, the TO phase shifters can be replaced with faster p-i-n phase shifters in our platform [22], with switching speed extended to the range of nanoseconds.

## Funding

This work was supported by the Academy of Finland Flagship Programme, Photonics Research and Innovation (PREIN, Decision No. 346545).

## Acknowledgement

The authors gratefully acknowledge Kauko Immonen (VTT) and Markku Kapulainen (VTT) for their contributions to the laboratory setup, and the microfabrication team at VTT for their technical support. The authors also thank Radu Lupoaie, Matti Meuronen, and Jarkko Pulli from Keysight Technologies for their assistance with equipment and laboratory configuration.

## Disclosures

The authors declare no conflicts of interest.

## Data availability

Data underlying the results presented in this paper are not publicly available at this time but may be obtained from the authors upon reasonable request.